%% file: main.tex
\documentclass[journal]{IEEEtran}

\usepackage{cite}

\ifCLASSINFOpdf
   \usepackage[pdftex]{graphicx}
   \graphicspath{{../pdf/}{../jpeg/}}
   \DeclareGraphicsExtensions{.pdf,.jpeg,.png}
\else
\fi

\UseRawInputEncoding
\usepackage{amsmath}
\usepackage{array}
\usepackage{standalone}
\usepackage{etoolbox}
\newtoggle{arXiv}
\newtoggle{bibrun}
\usepackage{newtxtext, newtxmath}
\usepackage{graphicx}
\usepackage{color, calc}
\usepackage{array, booktabs}
\newcolumntype{L}{>{$}l<{$}}
\newcolumntype{C}{>{$}c<{$}}
\newcolumntype{R}{>{$}r<{$}}
\usepackage{tikz, tikz-3dplot}
\usetikzlibrary{arrows, arrows.meta, calc, positioning}
\usetikzlibrary{decorations.pathmorphing}	
\tikzset{>=Stealth}
\usepackage{pgfplots}
\usepackage{siunitx, physics}
\usepackage{enumitem}
\setlist[description]{labelindent=0pt, leftmargin=\parindent, font=\normalfont\itshape}
\usepackage{url}
\usepackage{subfigure}
\usepackage{textcomp}
\usepackage{eurosym}

\begin{document}
%
\title{Peripheral Nerve Stimulation limits with fast narrow and broad-band pulses}

\author{\IEEEauthorblockN{
		Daniel~Grau-Ruiz\IEEEauthorrefmark{1}\IEEEauthorrefmark{2},
		Juan~P.~Rigla\IEEEauthorrefmark{2},
		Eduardo~Pall\'as\IEEEauthorrefmark{1},
		Jos\'e~M.~Algar\'{\i}n\IEEEauthorrefmark{1},
		Jose~Borreguero\IEEEauthorrefmark{1},
		Rub\'en~Bosch\IEEEauthorrefmark{1},
		Guillermo~Comazzi\IEEEauthorrefmark{1},
		Elena~D\'{\i}az-Caballero\IEEEauthorrefmark{2},
		Fernando~Galve\IEEEauthorrefmark{1},
		Carlos~Gramage\IEEEauthorrefmark{1},		
		Jos\'e~M.~Gonz\'alez\IEEEauthorrefmark{2},
		Rub\'en~Pellicer\IEEEauthorrefmark{1},
		Alfonso~Ríos\IEEEauthorrefmark{2},
		Jos\'e~M.~Benlloch\IEEEauthorrefmark{1}, and
		Joseba~Alonso\IEEEauthorrefmark{1}}
	
	\IEEEauthorblockA{\IEEEauthorrefmark{1}MRILab, Institute for Molecular Imaging and Instrumentation (i3M), Spanish National Research Council (CSIC) and Universitat Polit\`ecnica de Val\`encia (UPV), 46022 Valencia, Spain}\\
	\IEEEauthorblockA{\IEEEauthorrefmark{2}Tesoro Imaging S.L., 46022 Valencia, Spain}\\

\thanks{Corresponding author: J. Alonso (joseba.alonso@i3m.upv.es).}}


\maketitle

\begin{abstract}
Peripheral Nerve Stimulation (PNS) constrains the clinical performance of Magnetic Resonance and Particle Imaging (MRI and MPI) systems. Extensive magneto-stimulation studies have been carried out recently in the field of MPI, where typical operation frequencies range from single to tens of kilo-hertz. PNS literature is scarce for MRI in this regime, which can be relevant to small (low inductance) dedicated MRI setups, and where the resonant character of MPI coils prevents studies of broad-band excitation pulses. We have constructed an apparatus for PNS threshold determination on a subject's limb, capable of narrow and broad-band magnetic excitation with pulse characteristic times down to 40~$\boldsymbol{\upmu}$s. From a first set of measurements carried out on 51 volunteers, we observe that PNS limits coincide for sinusoidal (biphasic narrow-band) and triangular (biphasic broad-band) excitations, and are slightly lower for trapezoidal (monophasic broad-band) pulses. The observed dependence on pulse frequency/rise-time is compatible with traditional stimulation models where nervous responses are characterized by a rheobase and a chronaxie. We have also measured statistically significant correlations of PNS sensitivity with arm size and body weight, and no correlation with height or gender. As opposed to resonant systems, our setup allows the execution of arbitrarily short pulse trains. We have confirmed thresholds increase significantly as trains transition from tens to a few pulses also in these fast timescales. By changing the polarity of the coils in our setup, we also looked at the influence of the spatial distribution of magnetic field strength on PNS effects. We find that thresholds are higher in an approximately linearly inhomogeneous field (relevant to MRI) than in a rather homogeneous distribution (as in MPI). Finally, given the large intersubject variability of PNS sensitivity, we propose employing a versatile low-cost system (such as presented here) for fast offline determination of a subject's limits prior to medical scanning, and then using this information to boost clinical imaging while preserving the patient's safety.
\end{abstract}

 \ifCLASSOPTIONpeerreview
 \begin{center} \bfseries EDICS Category: 3-BBND \end{center}
 \fi
%
\IEEEpeerreviewmaketitle

\section{Introduction}
\IEEEPARstart{M}{agneto-stimulation} of the peripheral nervous system of patients is one amongst few safety concerns in clinical applications of Magnetic Resonance Imaging (MRI) and Magnetic Particle Imaging (MPI). In the former, Peripheral Nerve Stimulation (PNS) can take place when magnetic gradient fields used for spatial information encoding are pulsed on and off \cite{Irnich1995}; in the latter, when time varying (ac) magnetic fields excite the nanoparticles employed for background-free signal detection \cite{Saritas2013}. PNS is a physiological response to the presence of electric fields induced by time varying magnetic fields (as expected from the Maxwell-Faraday law of induction), which trigger action potentials in nerve or muscle fibers or bundles \cite{Reilly1989}. This is commonly perceived as a tingling or poking sensation, but can become uncomfortable or painful for strong fields, and even compromise the subject's health under extreme conditions \cite{Nyenhuis1994}.

The time varying magnetic fields employed in MRI and MPI setups are of different nature and serve different purposes. In MRI, magnetic gradient fields are most often ramped on or off, forming trapezoidal (broad-band spectrum) waveforms, whereas MPI employs spatially homogeneous fields which oscillate sinusoidally in time (narrow-band spectrum). However, the impact of magneto-stimulation effects on medical applications of both disciplines is notorious: it imposes stringent limits on diffusion weighted MRI \cite{Bammer2003}, as well as fast MRI pulse sequences including echo-planar imaging \cite{Stehling1991}, turbo spin echo \cite{Hennig1986} or steady-state free-precession techniques \cite{Carr1958}; and it has so far precluded whole-body MPI scanners \cite{Saritas2013}.

Numerous PNS experimental measurements and simulations have been conducted in the past decades \cite{Reilly1989,Schaefer2000,Zhang2003,Saritas2013,Davids2017,Klein2019}, but the interaction between dynamic magnetic fields and live tissues is complex and some aspects remain unclear \cite{Neufeld2016,Klein2020}. Some of the latest and most extensive PNS threshold measurements have been performed in the field of MPI \cite{Saritas2013,Demirel2017,Saritas2012}, which makes use of narrow-band excitation pulses with frequencies in the tens of kilo-hertz. In these studies, thresholds are measured against a number of parameters, including excitation frequency, duty cycle, body part, or coil dimensions. These measurements exploit the resonant nature of MPI coils to generate strong magnetic fields which boost scanner performance and facilitate magneto-stimulation studies. However, resonant circuits also constrain the rate at which magnetic pulses can be switched on and off (with time constants inversely proportional to the quality factor of the resonant circuit), they obstruct PNS threshold measurements against relevant variables (e.g. pulse train length), and they impede the use of broad-band pulses (relevant to MRI). On the other hand, MPI frequencies are fast compared to typical MRI timescales. Consequently, PNS threshold measurements for broad-band pulses switched in tens to hundreds of micro-seconds are scarce, even if this fast regime may be relevant for small, dedicated MRI systems, often based on permanent magnets and where field orientations may differ with respect to conventional systems \cite{Aksel2007,Ghazinoor2007,Rigla2020,Algarin2020,Cooley2020,OReilly2020}.

In this paper, we present an apparatus designed for PNS threshold determination on a subject's forearm, which allows for fast narrow and broad-band excitation pulses, and can be configured for different spatial magnetic field strength distributions (Sec.~\ref{sec:scanner}). We use this apparatus to extend previous studies to scenarios relevant to both MPI and MRI. From systematic measurements on 51 volunteers, we conclude that traditional magneto-stimulation models (presented in Sec.~\ref{sec:theory}) are accurate for excitation times down to tens of micro-seconds, for both narrow and broad-band pulses (Sec.~\ref{sec:protocol1}). Besides, we present in Sec.~\ref{sec:protocol2} studies performed on two volunteers, searching for qualitative patterns in the response to magneto-stimulation under different conditions. From these experiments we confirm that the position of the subject determines where and how PNS is experienced, we observe that the pulse train length can strongly influence PNS thresholds, and we conclude that PNS thresholds in our setup are significantly higher for an inhomogeneous field configuration (relevant to MRI) than for more homogeneous fields (as in MPI). Additionally, in Sec.~\ref{sec:concl} we investigate the possibility of using our apparatus for offline determination of PNS thresholds in order to boost the performance of imaging sequences in clinical MRI/MPI systems, based on the sensitivity of each individual.

\section{Theoretical framework}
\label{sec:theory}

According to the widespread model of electro-stimulation suggested in Ref.~\cite{Irnich1995}, a nervous response will be triggered if the time average of the electric field magnitude, $\tilde{E}\equiv\frac{1}{\tau}\int^\tau E \text{d}t$, surpasses a hyperbolic function of the time $\tau$ over which the electric field is active:
\begin{equation}\label{eq:Emean}
	\tilde{E}\geq E_\text{r}\left(1 + \frac{\tau_\text{c}}{\tau} \right),
\end{equation}
where $E_\text{r}$ is the so-called ``rheobase'', below which stimulation cannot take place, and $\tau_\text{c}$ is the ``chronaxie'', a time constant that determines how long it takes for the rheobase to be asymptotically reached. The link to magneto-stimulation is provided by the Maxwell-Faraday law:
\begin{equation}
	\oint\vec{E}\cdot\vec{\text{d}l} = \frac{\text{d}}{\text{d}t}\iint_S \vec{B}\cdot\vec{\text{d}S},
\end{equation}
where $\vec{\text{d}l}$ is a line element, $\vec{\text{d}S}$ is a surface element and $\vec{B}$ is the externally applied magnetic field. If we simplify the geometry of the exposed body part to be circular with radius $r$, the magnitude of the electric field along its perimeter is given by
\begin{equation}\label{eq:EandB}
	E=\kappa r \dot{B},
\end{equation}
where $\kappa\in[\frac{1}{2},1]$ is a form factor dependent on the orientation of the magnetic field vector \cite{Irnich1995}, and $\dot{B}$ denotes the time derivative of the magnetic field magnitude. From Eqs.~(\ref{eq:Emean}) and (\ref{eq:EandB}) we see that magnetostimulation will take place given that
\begin{equation}\label{eq:Bdot}
	\dot{B}	\geq \dot{B}_\text{r}\left(1 + \frac{\tau_\text{c}}{\tau} \right)\text{, with } \dot{B}_\text{r} = \frac{E_\text{r}}{\kappa r},
\end{equation}
or, alternatively, if the excursion of the magnetic field over a time $\tau$ obeys
\begin{equation}\label{eq:B}
	\Delta B(\tau)\geq \Delta B_\text{thr} = \Delta B_\text{r}\left(1 + \frac{\tau}{\tau_\text{c}} \right)\text{, with } \Delta B_\text{r}=\frac{E_\text{r}\tau_\text{c}}{\kappa r}=\dot{B}_\text{r}\tau_\text{c}.
\end{equation}
For a magnetic field oscillating with frequency $f$, this can be rewritten as:
\begin{equation}\label{eq:Bandf}
	\Delta B(f)\geq \Delta B_\text{thr} = \Delta B_\text{r}\left(1 + \frac{1}{2\tau_\text{c}f} \right),
\end{equation}
and $\Delta B$, $\Delta B_\text{thr}$ and $\Delta B_\text{r}$ are peak-to-peak excursions for $\tau=1/(2f)$. In the remainder of this paper, we will use mostly Eqs.~(\ref{eq:B}) and (\ref{eq:Bandf}). However, we find it helpful to make explicit the trivial relation with Eq.~(\ref{eq:Bdot}), which is how safety limits decreed by the International Electrotechnical Commission (IEC) are defined \cite{IEC2010}: $\dot{B}_\text{r,IEC}=20$~T/s (i.e. $E_\text{r,IEC}=2.2$~V/m for $r=22$~cm and $\kappa = 1/2$ in Eq.~(\ref{eq:EandB})) and $\tau_\text{c,IEC}=\SI{360}{\micro s}$, corresponding to $\Delta B_\text{r,IEC}=7.2$~mT. In order to compare our data analysis for triangular and sinusoidal (biphasic) waveforms with previous studies, we give $\Delta B_\text{r}$ as a peak-to-peak quantity. This does not apply for trapezoidal (monophasic) waveforms.

\section{Apparatus and control system}
\label{sec:scanner}

\begin{figure}
	\centering
	\includegraphics[width=\columnwidth]{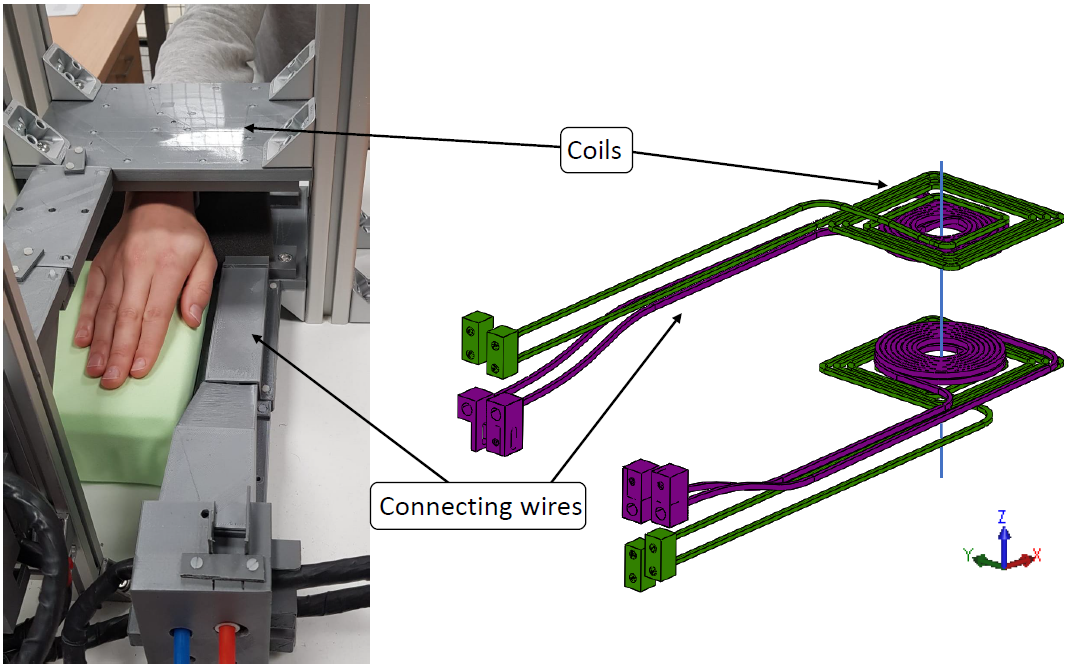}
	\caption{a) Photograph of PNS measurement setup. b) Coil design, consisting on two pairs of coils. We define the center of the coordinate system to be at the mid-point between the centers of the coil pairs (the $z$-axis is shown for further reference).}
	\label{fig:Setup}
\end{figure}

\begin{figure}
	\centering
	\includegraphics[width=\columnwidth]{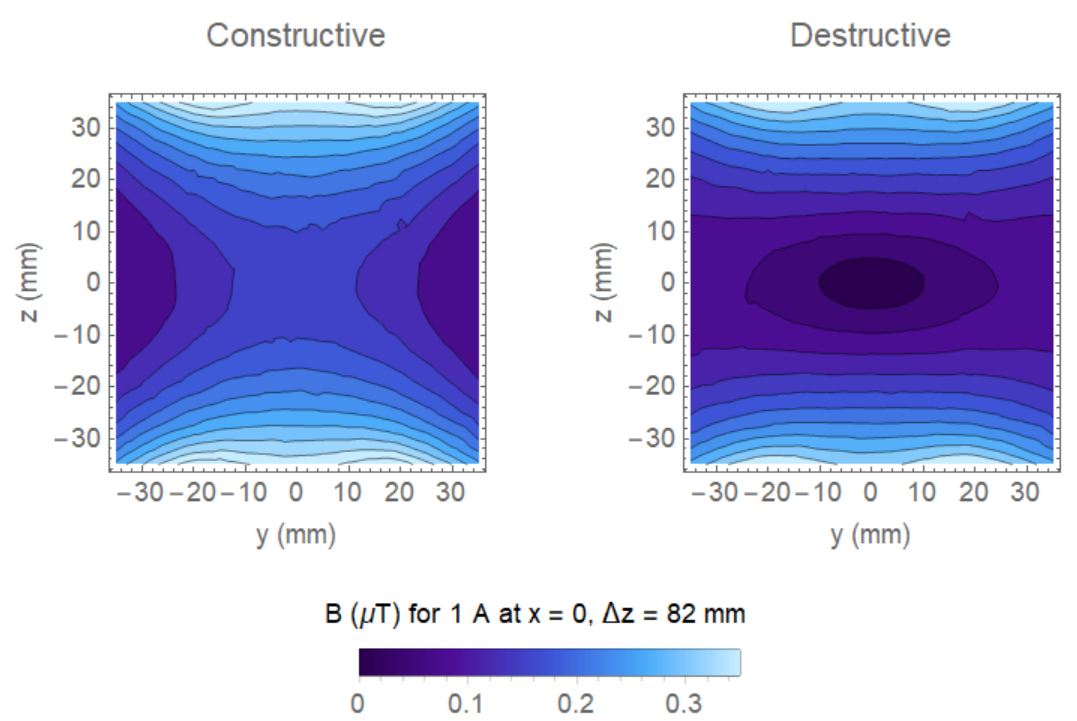}
	\caption{Magnetic field strength distributions for homogeneous (left) and inhomogeneous (right) configurations, simulated in Comsol for a current of 1~A running through the coils, and plotted for the $x=0$ plane, with $\Delta z=82$~mm. Structure along the contours is due to finite numeric precision in the simulations.}
	\label{fig:BFieldDist}
\end{figure}

All experiments reported in this paper have been performed in the PNS measurement setup in Fig.~\ref{fig:Setup}a). The top and bottom plates each enclose a pair of coils (Fig.~\ref{fig:Setup}b) which can generate dynamic magnetic fields for PNS experiments. The separation $\Delta z$ between plates can be manually adjusted from 0 to 200~mm, where we call $z$ the axis that goes through the center of both plates. All four coils are connected in series for a total inductance of $\approx\SI{43}{\micro H}$. The coil system can be configured to generate a rather homogeneous magnetic field strength distribution (constructive configuration), with a maximum at the center of the $z=0$ plane, or one which varies approximately linearly along the $z$-axis (destructive configuration), which nulls at the center. Figure~\ref{fig:BFieldDist} shows the field maps for both configurations for $\Delta z \approx 82$~mm. The field is mostly transverse to the forearm cross-section, justifying the use of $\kappa=1$ to calculate electric field amplitudes in the below analysis.

The coils are made of hollow-tube Oxygen-Free High-Conductivity (OFHC) copper to constrain ohmic losses and facilitate heat removal with chilled water. Continuous operation of the system is possible with $\approx\SI{2.6}{l/min}$ of cooling water, for a pressure drop of $\approx\SI{0.4}{bar}$ and a thermal jump $<\SI{2}{K}$ between the water inlet and outlet. 

The control system and Graphical User Interface (GUI) have been programmed in LabVIEW (National Instruments). Here, the experimenter can choose between different waveforms (sinusoidal, triangular or trapezoidal) and pulse train characteristics (train length, frequency, ramp times, etc.). The software is programmed to scan the pulse train amplitude for a given waveform and set of parameters, for which the volunteer is asked whether magneto-stimulation has taken place after every iteration (pulse train with a given amplitude). The experimenter also records whether the subject has been stimulated in the GUI, which stores the data in a folder and file structure that we use for data analysis with Mathematica (Wolfram). 

A digital-to-analog converter (National Instruments VirtualBench VB-8012) is serially connected to the control computer to read in the designed dynamics and generate a low-voltage analog waveform. This is amplified to up to $\pm 400$~A using a single channel of a high power gradient amplifier (International Electric Co. GPA-400-750). With a maximum voltage of 750~V and current of 400~A, slew rates  $>\SI{15}{A/\micro s}$ are possible on our load. For $\Delta z \approx 82$~mm in the constructive configuration, this translates into a maximal $\dot{B}\approx 5000$~T/s at $z=\pm30$~mm (characteristic forearm radius). The load inductance and parasitic capacitances lead to a non-linear dependence of the amplitude on the pulse train time characteristics. All the below results are already corrected for these effects, which we previously calibrate by comparing the amplifier outputs against the nominal input values for all the employed waveforms and pulse train parameters.

\section{Quantitative study on large population}
\label{sec:protocol1}

The goal of this study is to determine magneto-stimulation thresholds for narrow and broad-band magnetic pulse trains, on a sizable population and for characteristic excitation times $\tau$ going from \SI{250}{\micro s} down to \SI{42}{\micro s}, corresponding to frequencies $f=1/(2\tau)$ between 2 and 12~kHz for oscillatory pulses. We ran the same protocol on 51 volunteers, following procedures approved by ethical committees at La Fe Hospital in Valencia, the Spanish National Research Council in Madrid, and the European Commission in Brussels. All measurements in this study are for a constructive field distribution (see Sec.~\ref{sec:scanner}) and $\Delta z \approx 82$~mm to ensure that all subjects are exposed to congruent magnetic field dynamics.

\begin{figure}
	\centering
	\includegraphics[width=\columnwidth]{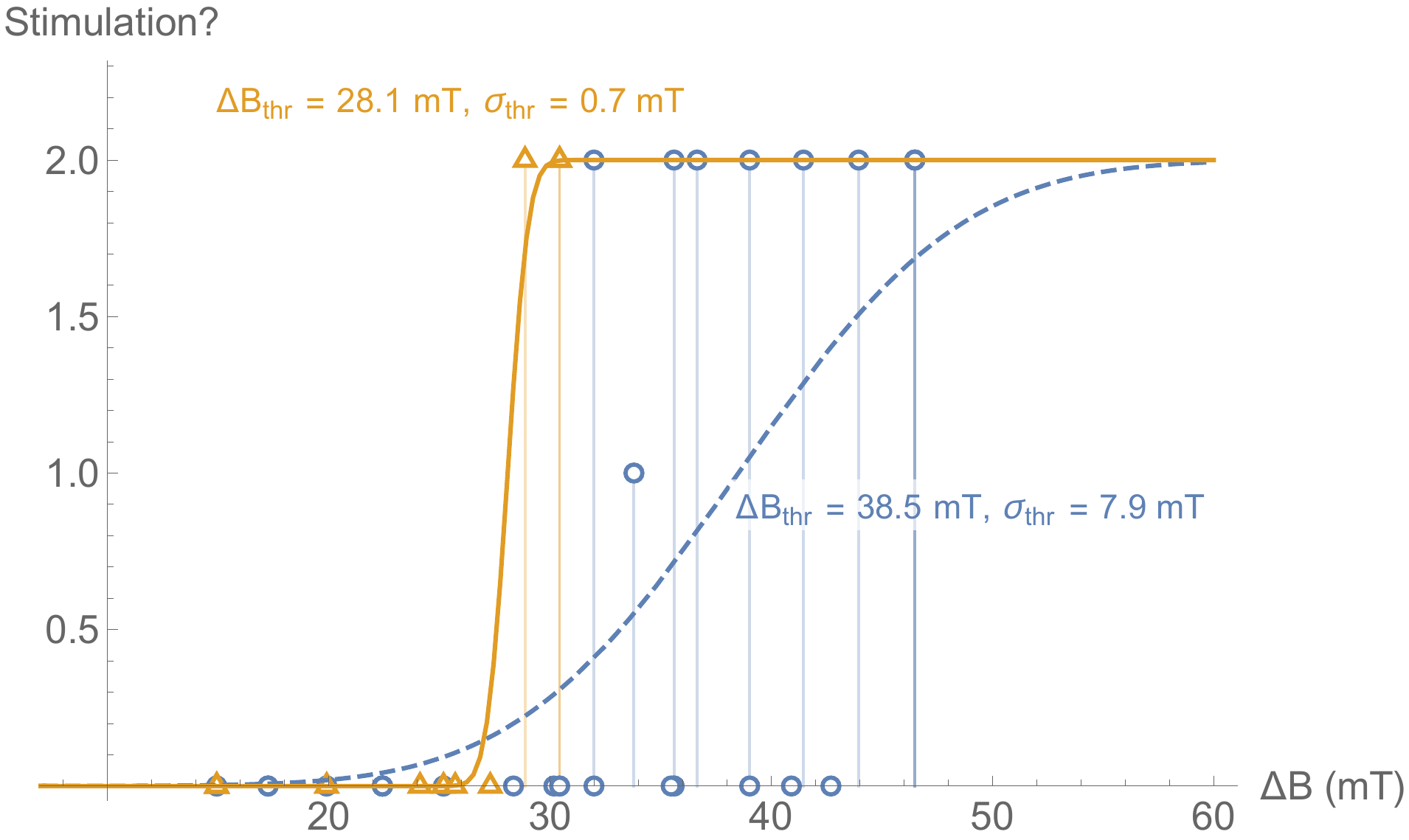}
	\caption{Scan results for a triangular waveform at 5~kHz from two different volunteers, one with a clear threshold (orange) and one noisier measurement (blue). The data are fit to the model in Eq.~(\ref{eq:model}) to determine the PNS thresholds and uncertainties (see text).}
	\label{fig:SampleScanFit}
\end{figure}

For these experiments, subjects introduced their right forearm, positioned such that the $z$-axis crossed it at a distance $\approx 5$~cm from the wrist towards the elbow (see Fig.~\ref{fig:Setup}a). After a short training session to familiarize the subject with the setup, the dynamics of the experiment and, especially, their particular perception of PNS, we ran systematic scans with trains of 1,000 pulses for three different waveforms: sinusoidal (narrow-band), and triangular and trapezoidal (broad-band) pulse shapes. For a given waveform and time configuration (i.e. frequency for sinusoidal and triangular pulse trains, ramp time for trapezoids), we first carried out a coarse amplitude scan (with a typical step of 25~A, corresponding to field jumps of $\approx 3$~mT at the origin), which stopped once the subject reported stimulation. For every pulse train (i.e. amplitude) in the scan, we recorded a ``0'' if there was no stimulation, a ``1'' when the subject was unsure, and a ``2'' for confirmed stimulation. After the coarse scan, we swept randomly a small region below the first ``2'' with finer resolution (typically 8~A, i.e. $\approx 1$~mT) and including a few ``placebo'' (low intensity) pulses. This allows a more accurate determination of the stimulation threshold for the selected waveform and time parameters by fitting an error function (erf) to the data points, including both the coarse and fine sweeps. In particular, we fit the model
\begin{equation}\label{eq:model}
	1 + \text{erf}\left( \frac{\Delta B-\Delta B_\text{thr}}{\sigma_\text{thr}\sqrt{2}} \right),
\end{equation}
where $\text{erf}(x)=\frac{2}{\sqrt{\pi}}\int_0^x\text{e}^{-t^2}\text{d}t$ is the error function, $\Delta B$ is the excursion in magnetic field amplitude for a given waveform and time parameters, and $\Delta B_\text{thr}$ is the PNS threshold, to which we assign an error bar of size $\sigma_\text{thr}$ (constrained to be larger than thrice the fine step). For illustration purposes, Fig.~\ref{fig:SampleScanFit} shows the data and fits for scans from two different volunteers for the same waveform and time parameters. The field strength given corresponds to the center of coordinates.

\begin{figure*}
	\centering
	\includegraphics[width=2\columnwidth]{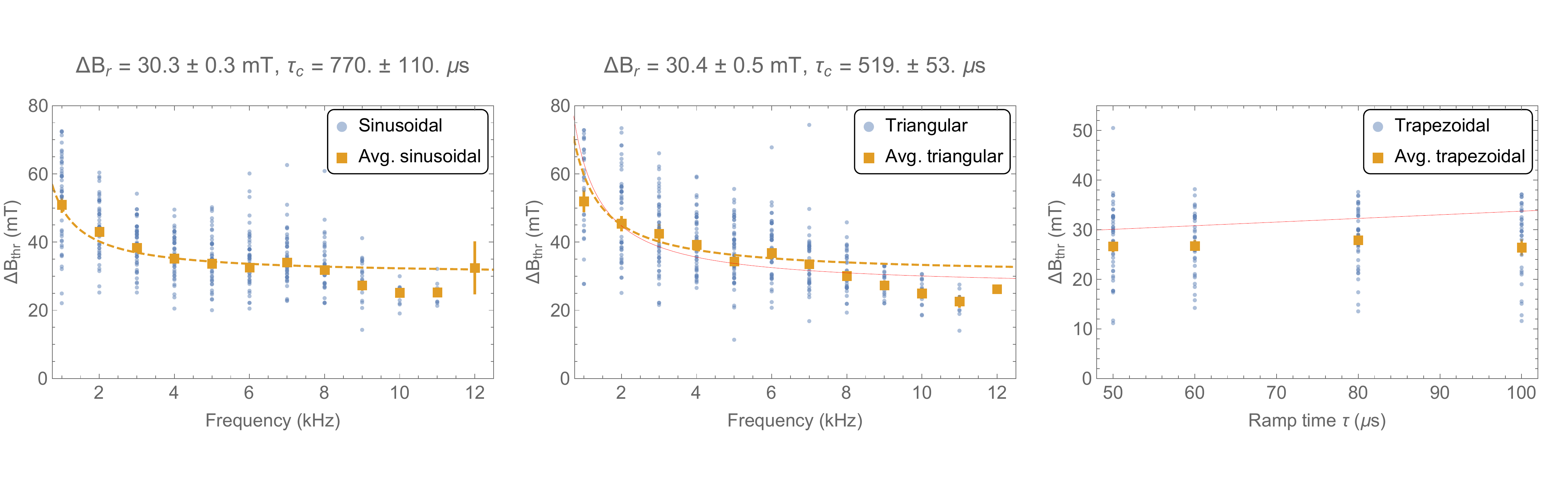}
	\caption{Experimental PNS threshold determination of sinusoidal (left), triangular (middle) and trapezoidal (right) trains of 1,000 pulses on 51 volunteers. Blue circles are semi-transparent and represent the threshold for every individual, so darker points represent multiple overlapping measurements. Orange squares denote the mean for a given waveform and time parameters, weighted by the inverse of the $\sigma_\text{thr}$ values obtained from fitting to error functions (see text). The averaged values are plotted with error bars, but most are hidden behind the markers. The dashed orange lines are fits to Eq.~(\ref{eq:Bandf}), with $\Delta B_\text{r} = 30.3\pm 0.3$~mT ($30.4\pm 0.5$~mT) and $\tau_\text{c} = 770\pm\SI{110}{\micro s}$ ($519\pm\SI{53}{\micro s}$) for the sinusoidal (triangular) scans. The red curves in the scans with linear magnetic ramps correspond to the IEC default thresholds, linearly extrapolated from $r=22$~cm (relevant to whole body systems) to $r=3$~cm (typical forearm size), and with $\kappa=1$ instead of 1/2 (see text). Note that $\Delta B_\text{thr}$ in the trapezoidal plot corresponds to the amplitude of the monopolar pulses, whereas it is a peak-to-peak amplitude in the sinusoidal and triangular plots.}
	\label{fig:AllVolunteers}
\end{figure*}

The plots in Fig.~\ref{fig:AllVolunteers} are the main result for this set of experiments. The left plot shows the average (weighted by $1/\sigma_\text{thr}$ in Eq.~(\ref{eq:model})) of the PNS threshold ($\Delta B_\text{thr}$) for sinusoidal trains of 1,000 pulses as a function of their frequency. The field strength again corresponds to the origin. All individual threshold measurements are included (small blue points) to highlight the large spread of magneto-stimulation limits within the scanned  population. Error bars in the averaged data are calculated as the standard deviation of the mean. We also fit each dataset to Eq.~(\ref{eq:Bandf}) to determine a rheobase and chronaxie time for every volunteer, for sinusoidal pulse trains. The curve corresponding to the mean values ($\Delta B_\text{r} = 30.3\pm 0.3$~mT, $\tau_\text{c} = 770\pm\SI{110}{\micro s}$) is also plotted in the figure. The trend follows closely the model in Eq.~(\ref{eq:Bandf}), but the determined chronaxie is significantly longer than the one reported in Ref.~\cite{Saritas2013} for a similar experiment on 20 volunteers ($284\pm\SI{67}{\micro s}$). This is likely due to differences in the magnetic field distributions: the solenoids in Ref.~\cite{Saritas2013} provide homogeneous longitudinal fields over a large field of view, whereas our coils generate a less homogeneous transverse field (Fig.~\ref{fig:BFieldDist}). High frequency data appear systematically below the fit, as expected from the fact that only particularly sensitive volunteers were stimulated at the maximum currents available at these frequencies (note the lower abundance of blue points).

\begin{figure}
	\centering
	\includegraphics[width=0.8\columnwidth]{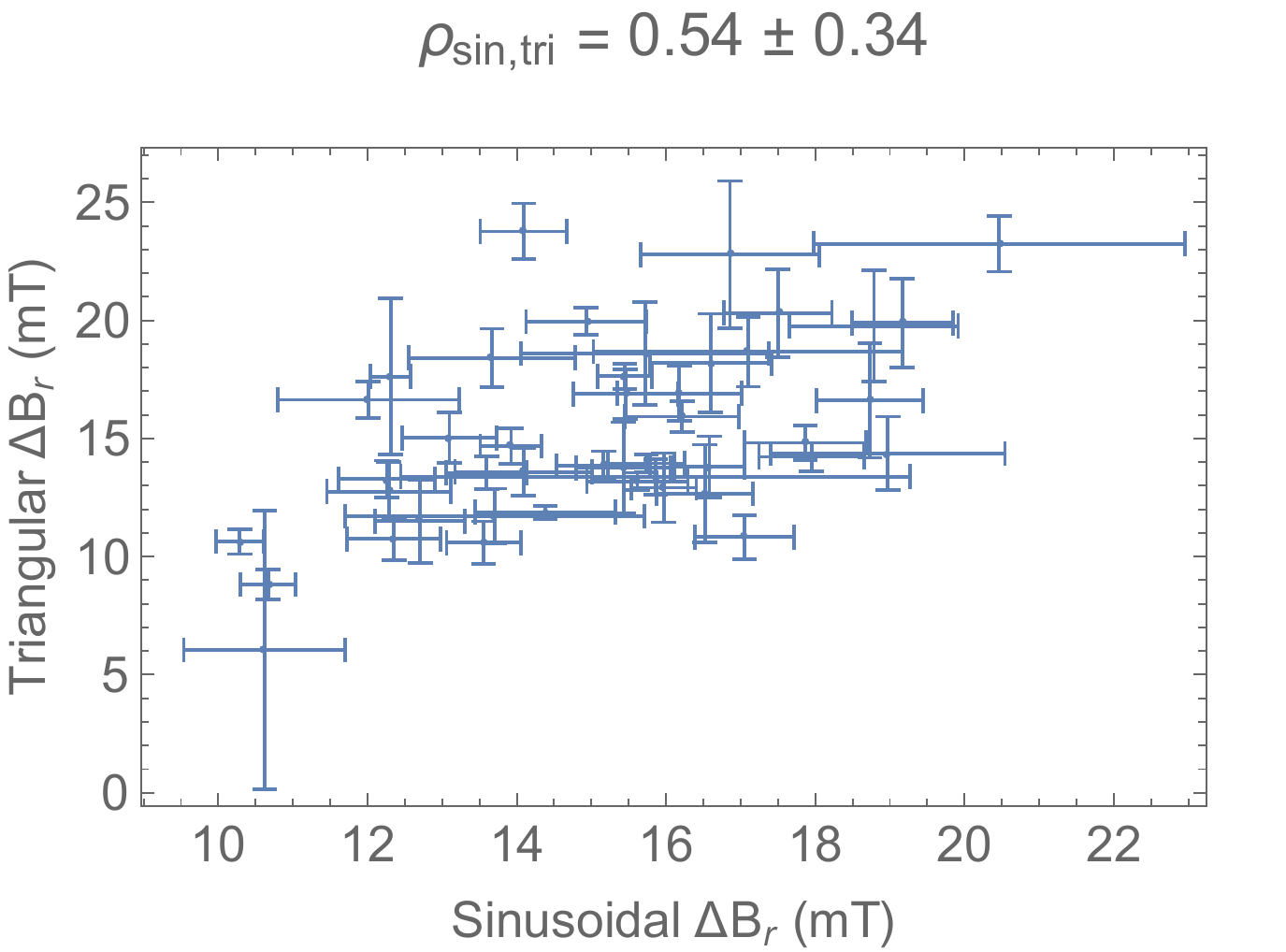}
	\caption{Sinusoidal and triangular rheobase $\Delta B_\text{r}$ for the 51 volunteers in the quantitative study (Sec.~\ref{sec:protocol1}). The Pearson correlation coefficient is $\rho_\text{sin,tri} = 0.54 \pm 0.34$.}
	\label{fig:Corrs2}
\end{figure}

Aside from the above, our results include broad-band excitations. The middle and right plots show measurements for triangular and trapezoidal waveforms respectively. Here we include also the IEC default limits, linearly extrapolated from $r=22$~cm (relevant to whole body systems) to $r=3$~cm (typical forearm size), and assuming $\kappa = 1$ instead of 1/2, so $\Delta B_\text{r,IEC}=26.4$~mT. The thresholds measured for triangular pulse trains follow a similar pattern to the sinusoidal case, with a consistent mean rheobase $\Delta B_\text{r} = 30.4\pm 0.5$~mT and a chronaxie time $\tau_\text{c} = 519\pm\SI{53}{\micro s}$. From these quantities and Eq.~(\ref{eq:B}), we estimate an electric field rheobase $E_\text{r} = 1.8\pm 0.2$~V/m, not far from the IEC reference value ($E_\text{r,IEC} = 2.2$~V/m, \cite{IEC2010}). The individual rheobases determined for sinusoidal and triangular excitations correlate strongly, with a Pearson coefficient (covariance divided by the product of the standard deviations) $\rho_\text{sin,tri} = 0.54 \pm 0.34$ (Fig.~\ref{fig:Corrs2}). For the trapezoidal case we measured four different rise/fall times between 50 and $\SI{100}{\micro s}$ (significantly shorter than the chronaxie). Here, a linear regression to Eq.~(\ref{eq:B}) would be under-constrained and would not deliver meaningful fit parameters, but the data suggest $\Delta B_\text{r} < 30$~mT. This is slightly below the values obtained for biphasic waveforms, as expected for monophasic pulses \cite{Reilly1985}.

From the above sets of results we conclude that traditional magneto-stimulation models (Eqs.~(\ref{eq:B}) and (\ref{eq:Bandf})) accurately describe our findings, and we observe that the measured thresholds are consistent with the extrapolated IEC limits. On the other hand, the large inter-subject variability measured in this and other studies motivates the potential relevance of offline personalized PNS threshold measurements, which we discuss in Sec.~\ref{sec:concl}.

\begin{figure}
	\centering
	\includegraphics[width=0.9\columnwidth]{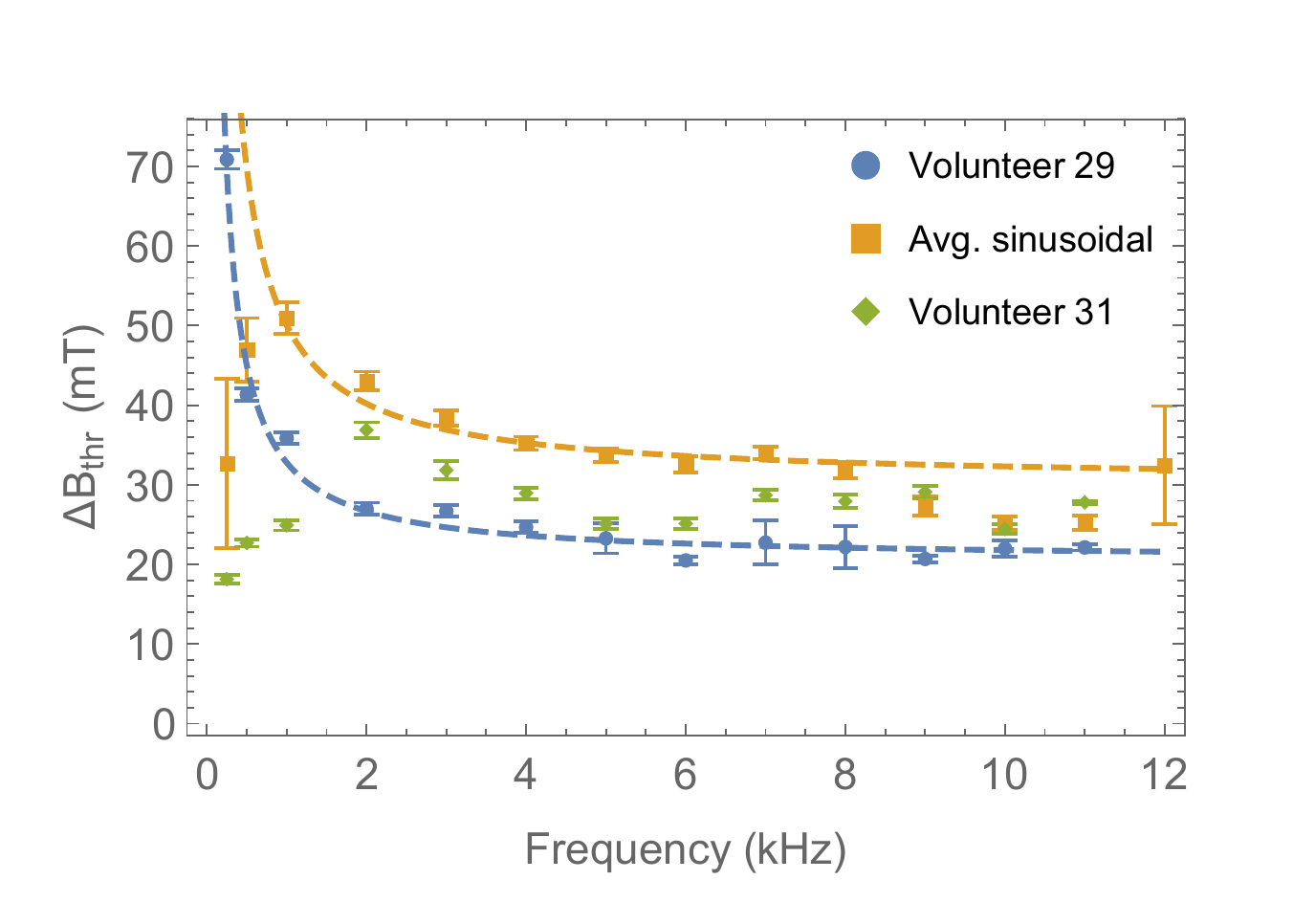}
	\caption{Experimental PNS threshold determination of sinusoidal trains of 1,000 pulses on two volunteers: one who likely distinguished magnet-stimulation from mechanical vibration (blue circles), and one who likely did not (green diamonds). The average sinusoidal threshold (for all 51 volunteers) is also shown (orange squares). The blue dashed line is a fits to Eq.~(\ref{eq:Bandf}), yielding $\Delta B_\text{r} = 20.6\pm 0.6$~mT and $\tau_\text{c} = 840 \pm \SI{60}{\micro s}$. The dashed orange line is the same as in Fig.~\ref{fig:AllVolunteers} (left), with $\Delta B_\text{r} = 30.3\pm 0.3$~mT and $\tau_\text{c} = 770 \pm \SI{110}{\micro s}$, where frequencies below 1~kHz are excluded from the fit.}
	\label{fig:LowFreq}
\end{figure}

Our protocol includes also measurements at frequencies below 1~kHz for sinusoidal and triangular waveforms (specifically at 250 and 500~Hz). These are not shown in Fig.~\ref{fig:AllVolunteers} because, in this regime, magnetic forces induce strong mechanical vibrations on the coils and surrounding structure, and some volunteers could not discern magneto-stimulation from vibration effects. As a result, low frequency measurements are fewer and qualitatively different for volunteers that distinguished one from the other (for whom the hyperbolic law remains accurate) and those that did not (with lower thresholds at the lowest frequencies, where mechanical vibrations are strongest). This is illustrated in Fig.~\ref{fig:LowFreq}: volunteer~29 proved to be highly sensitive to PNS, featuring thresholds well below average ($\Delta B_\text{r} = 20.6\pm 0.6$~mT, $\tau_\text{c} = 840\pm\SI{60}{\micro s}$), and they seemed not to be misled by mechanical vibrations (as opposed to volunteer~31).

\begin{figure*}
	\centering
	\includegraphics[width=2\columnwidth]{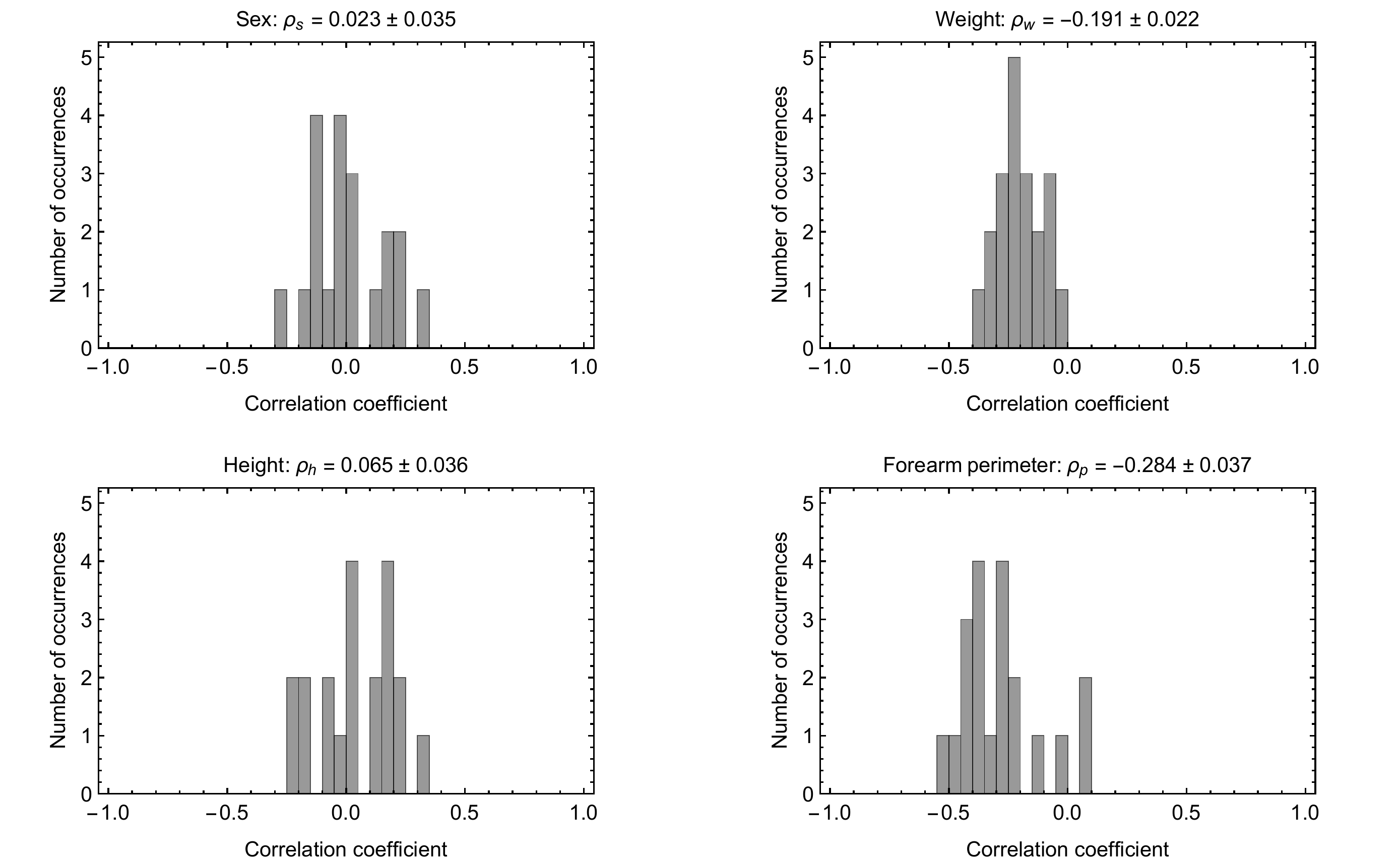}
	\caption{Histograms of Pearson correlation coefficients ($\rho$) between PNS thresholds (for all waveforms and time parameters, excluding frequencies below 1~kHz and above 8~kHz) and four physiological details from the volunteers: sex, weight, height and forearm perimeter.}
	\label{fig:Corrs}
\end{figure*}

Finally, as part of this study, we collected some personal details from the participants to establish correlations between their PNS sensitivity and physiological characteristics. In Fig.~\ref{fig:Corrs} we show correlations with the volunteers' sex, weight, height and forearm perimeter. These are first calculated individually for every combination of waveform and time configuration, as Pearson correlation coefficients. Each plot in Fig.~\ref{fig:Corrs} is a histogram for these coefficients, with their mean and standard deviation shown above. From these results we conclude there is no obvious correlation with gender or height. Previous studies performed on whole-body systems have found strong correlations with gender (see e.g. \cite{Faber2003}), but our results suggest that this is probably due more to body size than different perception thresholds. On the other hand, we observe a statistically significant negative correlation with body mass and forearm perimeter. This is consistent with previous experiments \cite{Saritas2013} and the explicit dependence of $\Delta B_\text{r}$ on $r$ in Eq.~(\ref{eq:B}).

\section{Qualitative studies on single volunteers}
\label{sec:protocol2}

Here we describe three additional studies that extend the quantitative results presented in the previous section with more qualitative observations about magneto-stimulation patterns under different circumstances.

\begin{figure}
	\centering
	\includegraphics[width=0.95\columnwidth]{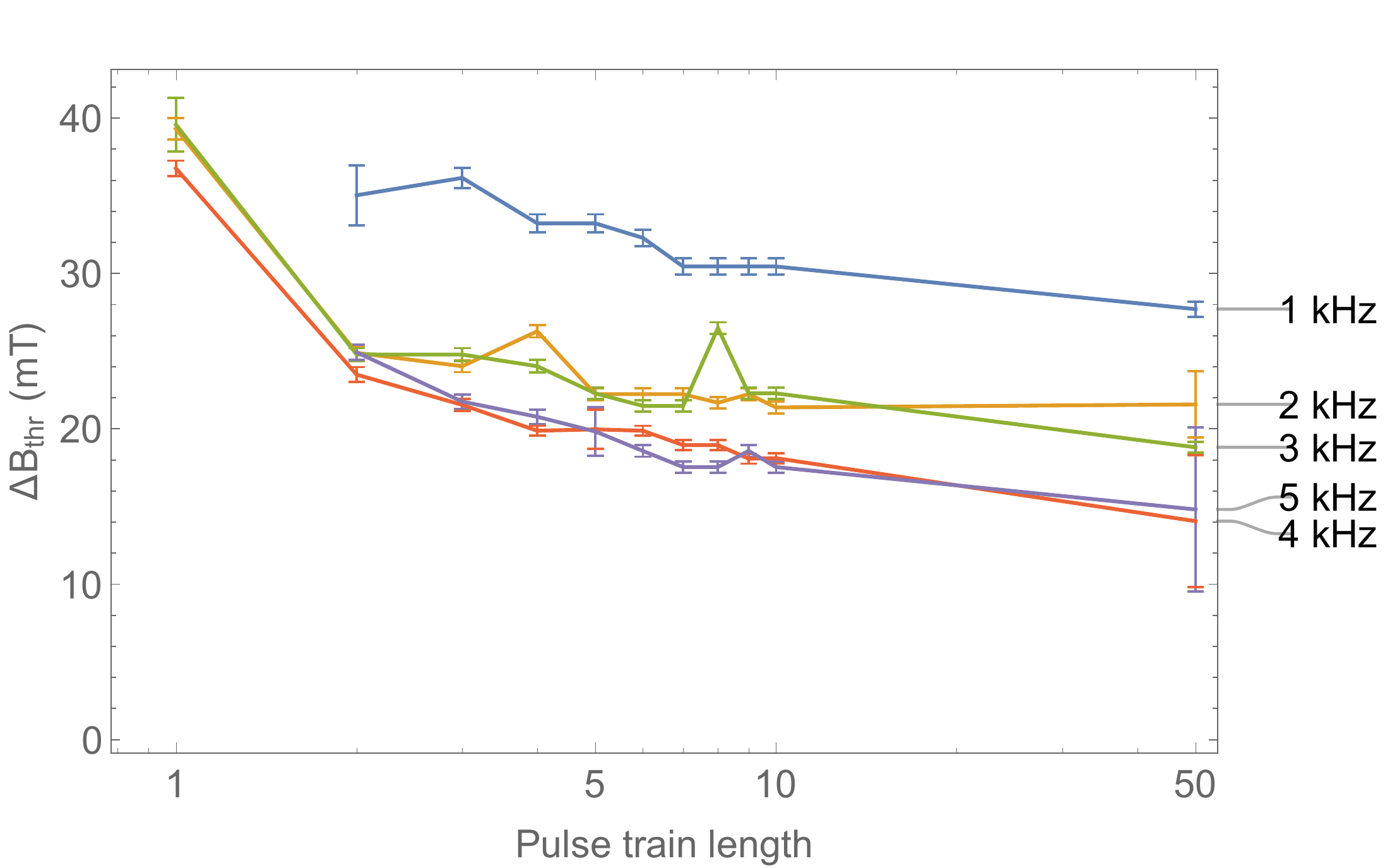}
	\caption{PNS thresholds for volunteer 8 when subject to trains of sinusoidal pulses from 1 to 5~kHz, with varying pulse train lengths. As in Fig.~\ref{fig:AllVolunteers}, $\Delta B_\text{thr}$ is the field excursion at $x=y=z=0$ in the constructive field configuration. Note that at 1~kHz the PNS threshold is higher than the maximum field excursion we can generate.}
	\label{fig:PulseTrainLength}
\end{figure}

Firstly, we investigate the influence of the pulse train length on the reported PNS threshold. To this end, we exposed volunteer 8 to trains of 1 to 50 sinusoidal pulses, with frequencies between 1 and 5~kHz. PNS thresholds and their uncertainties are determined in the same manner as described in Sec.~\ref{sec:protocol1}, and the measured data points are shown in Fig.~\ref{fig:PulseTrainLength}. Beyond the fact that thresholds decrease with increasing frequency, in agreement with the results in Fig.~\ref{fig:AllVolunteers}, a second trend is clear: they increase for short pulse trains. This effect was observed for whole-body systems at lower frequencies ($\approx 1$~kHz) already decades ago \cite{Budinger1991}, and has been conjectured to be behind deviations from the model in Eqs.~(\ref{eq:Bdot})-(\ref{eq:Bandf}) observed for very high frequencies ($>50$~kHz, \cite{12Weinberg}) but not reproducible in other MPI systems \cite{Saritas2013,Schmale2013}. Note that the systems in Refs.~\cite{12Weinberg,Saritas2013,Schmale2013} are all resonant (narrow-band), and therefore limited in their ability to investigate the influence of the pulse train length on the magneto-stimulation thresholds.

\begin{figure}
	\centering
	\includegraphics[width=0.95\columnwidth]{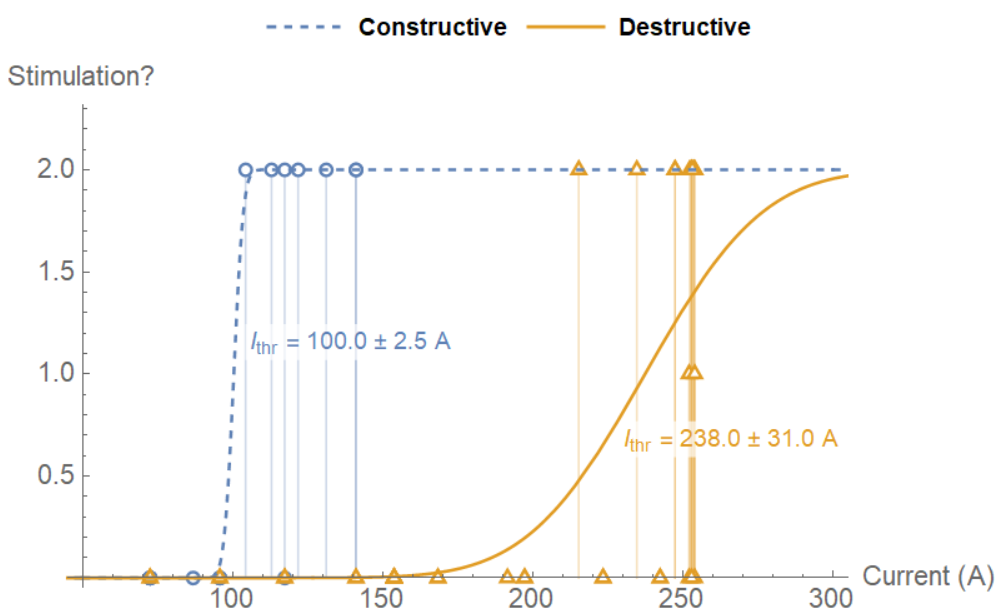}
	\caption{Scan results for volunteer 29 when subject to a train of 1,000 sinusoidal pulses at 5~kHz in the constructive (dashed blue) and destructive (solid orange) magnetic field configuration in Fig.~\ref{fig:BFieldDist}.}
	\label{fig:ConsDes}
\end{figure}

Secondly, we measured the PNS thresholds of volunteer 29 for a given waveform (sinusoidal trains of 1,000 pulses at 5~kHz) for the constructive and destructive field configurations. Figure~\ref{fig:ConsDes} shows that the spatial field distribution plays a critical role in magneto-stimulation, with threshold intensities twice as high in the destructive than in the constructive configuration, even if field intensities at $r=30$~mm are comparable for a given coil current (see Fig.~\ref{fig:BFieldDist}).

Lastly, the arm position is expected to influence the electric fields induced in the body and therefore which nerve fibers are triggered, as well as the strength of their response. To test this, we ran systematic scans on volunteer 29. Here, we also used sinusoidal trains of 1,000 pulses at 5~kHz, with a peak current of 125~A (slightly above the threshold reported by the subject in the default position) in the constructive field configuration. The volunteer started with their elbow at the center of the coils. As they pulled the arm out in six steps until the wrist was at the coil center, they went from no stimulation, to stimulation at the tip of the thumb, to stimulation on both thumb and index (suggesting radial and median nerve stimulation), with intensity fading as the wrist approached the coil center. We then ran a similar test with the hand palm facing upwards, and again a change in perception location and intensity was reported, in this case taking place at the little, middle and ring fingers (suggesting ulnar and median nerve stimulation). The most intense effect was with the palm facing left and the wrist close to the coil center, where the volunteer felt pain in the thumb, index and middle fingers.

\section{Conclusion}
\label{sec:concl}

Determining individual PNS thresholds can be critical in medical applications of MRI and MPI. For instance, the performance of MRI gradients determines the spatial resolution and acquisition times in fast imaging sequences. This is typically constrained by hardware limitations for short or long ramp times, and by PNS in intermediate regimes \cite{Chronik2001}. The main regulatory bodies in matters of MRI medical safety base their mandates on directive 60601-2-33:2010 from the International Electrotechnical Commission (IEC, \cite{IEC2010}), which relies on Spatially Extended Nonlinear Node models (SENN,\cite{Reilly1985}). This requires gradient systems to function in a ``normal operating mode'' bounded by parameters which can take either default values defined in the directive (see Sec.~\ref{sec:theory}), or values determined experimentally as 80~\% of the gradient settings for which half of the subjects are magneto-stimulated. A drawback of this approach is the use of average stimulation thresholds, since the simple mean does not reflect the large inter-subject variability that measurements reveal: PNS thresholds can vary by large factors ($>\times 3$) depending on physiognomy (see Sec.~\ref{sec:protocol1}). Hence, the IEC directive prevents stimulation for a majority of patients, but at the cost of infra-utilizing the scanner in cases where subjects could tolerate faster sequences.

Ideally, this would be circumvented by performing individual measurements of the subject's PNS thresholds and then adapting the pulse sequence parameters to maximize the system's performance. However, this is severely impractical due to the reduction of useful time it would impose on the scanner.

Realistic simulations of the physiological response of the peripheral nervous system to dynamic magnetic fields are an option to gain insight about the relevant interactions and processes \cite{Davids2017,Klein2019,Davids2019}. Currently, however, these can hardly be used to estimate the exact PNS thresholds for a given patient. Indeed, the above tests point out the extreme challenge of predicting PNS thresholds in clinically relevant settings: magneto-stimulation is triggered when the second spatial derivative of the induced electric fields along the nerve fibers is large enough to deplete the charges and activate the polarization pumps, and this is highly dependent on the subject's physiognomy and exact position with respect to the generated fields.

Here we propose to use a low cost setup for offline determination of the sensitivity of every individual to magneto-stimulation effects, then use this information to set scanner parameters to operate at its highest possible performance, while avoiding PNS. The system in Fig.~\ref{fig:Setup} is an inexpensive (< 50~k\euro) candidate, allowing for fast (< 2~min) estimation of a subject's individual sensitivity, stimulation with arbitrary waveforms (narrow and broad-band) down to $\tau<\SI{50}{\micro s}$, and where PNS effects can be studied in a rather homogeneous setting (relevant to MPI) as well as a ``gradient'' configuration (approximately linear inhomogeneity, relevant to MRI).

\section*{Contributions}

Experimental data were taken by DGR, using an apparatus built by DGR, JPR, EP, JMG, CG and JA, with contributions from JB, RB, GC and EDC. Data analysis performed by JA and DGR, with input from JMA, FG and RP. The paper was written by JA and DGR with input from all authors. Experiments conceived by JMB, JA and AR.


\appendices


\section*{Acknowledgment}

We thank all 51 anonymous volunteers for their participation in these studies, and Manuel Murbach for discussions. This work was supported by the European Commission under grant 737180 (FET-Open: HistoMRI) and Ministerio de Ciencia e Innovaci\'on of Spain for research grant PID2019-111436RB-C21. Action co-financed by the European Union through the Programa Operativo del Fondo Europeo de Desarrollo Regional (FEDER) of the Comunitat Valenciana 2014-2020 (IDIFEDER/2018/022).

\section*{Ethical statement}
All experiments were carried out following Spanish regulations and under the ethical consent from La Fe Hospital in Valencia (IIS-F-PG-22-02, agreement number 2019-139-1), approved by the Spanish National Research Council and the European Commission in the scope of research grant number 737180 (FET-Open: HistoMRI). Informed consent was obtained from all participants prior to study commencement.
	
\ifCLASSOPTIONcaptionsoff
  \newpage
\fi

\input{main.bbl}

\end{document}